\documentclass[USenglish,twocolumn]{article}

\ifx\directlua\undefined\ifx\XeTeXcharclass\undefined
  \usepackage[utf8]{inputenc}                           
  \else\RequirePackage[no-math]{fontspec}[2017/03/31]\fi 
  \else\RequirePackage[no-math]{fontspec}[2017/03/31]\fi 
\usepackage[sort&compress,square,numbers]{natbib}
\usepackage[big,online]{dgruyter}

\theoremstyle{dgthm}

\theoremstyle{dgdef}

\begin{document}

\articletype{Research Article}

\author*[1]{Willie J. Padilla}
\author[2]{Yang Deng}
\author[3]{Omar Khatib}
\author[4]{Vahid Tarokh} 
\affil[1]{Pratt School of Engineering, Duke University, Durham, North Carolina,27705, USA, willie.padilla@duke.edu; https://orcid.org/0000-0001-7734-8847}
\affil[2]{Pratt School of Engineering, Duke University, Durham, North Carolina,27705, USA, omar.khatib@duke.edu; https://orcid.org/0000-0002-0721-9684}
\affil[3]{Pratt School of Engineering, Duke University, Durham, North Carolina,27705, USA, yang.deng@duke.edu; https://orcid.org/0000-0003-2249-6556}
\affil[4]{Pratt School of Engineering, Duke University, Durham, North Carolina,27705, USA, vahid.tarokh@duke.edu}
\title{Fundamental absorption bandwidth to thickness limit for transparent homogeneous layers}
\runningtitle{Fundamental absorption bandwidth to thickness limit}
\abstract{Past work has considered the analytic properties of the reflection coefficient for a metal-backed slab. The primary result established a fundamental relationship for the minimal layer thickness to bandwidth ratio achievable for an absorber. There has yet to be establishment of a similar relationship for non metal-backed layers, and here we present the universal result based on the Kramers-Kronig relations. Our theory is validated with transfer matrix calculations of homogeneous materials, and full-wave numerical simulations of electromagnetic metamaterials. Our results place more general fundamental limits on absorbers and thus will be important for both fundamental and applied studies.}
\journalname{Nanophotonics}
\journalyear{2023}
\journalvolume{aop}

\maketitle

\section{Introduction}
The quantities of absorption and emission are fundamental processes related to the energy contribution of radiative heat transfer. Kirchhoff’s law of thermal radiation derives from the second law of thermodynamics and indicates that the emissivity ($\mathcal{E}$) and absorptivity (A) of a surface are identical in thermal equilibrium over spectral, directional, and solid angle degrees of freedom. \cite{Howell2020} Therefore, for any system operating in thermal equilibrium an understanding of the fundamental limits placed on absorption is of paramount importance. Physical bounds offer insights into the peak performance an absorber can achieve, given a specific material type and size. The potential of universally derived bounds, irrespective of specific materials, can be significant in developing and analyzing various absorber types. These bounds can further serve as benchmarks for evaluating the performance and stopping criteria in optimization of absorber bandwidth. \cite{Ye2013,Radi2015,Glybovski2016,Jung2020,Wang2023} Only after gaining a thorough understanding of the fundamental limits of absorption can more complex scenarios in radiative transfer be explored.

One fundamental limit established the relationship between absorber thickness and absorption bandwidth, and is valid for any single-layer (or multi-layer) metal backed structure. An assumption is that the reflection coefficient can be described in the long wavelength limit, keeping only the first-order term in $\lambda^{-1}$, which is given as, 

\begin{equation}
   \tilde{r}(\lambda)|_{\lambda\rightarrow\infty}=-1+i\frac{4\pi\mu d}{\lambda_0}
\label{ras}
\end{equation}

\noindent Using the reflection coefficient given in Eq. \ref{ras}, a Kramers-Kronig analysis shows that the limit for a metal-backed absorber of thickness ($d_{RL}$) is given by, \cite{Rozanov2000}

\begin{equation}
    d_{RL}\geq\frac{1}{2\pi^2\mu_s}\left|\int_{0}^{\infty} \ln{|\tilde{r}(\lambda)|} \,d\lambda \right| \equiv d_{R}
    \label{Rozanov_Lim}
\end{equation}

\noindent where $\mu_s = \textit{Re}\{\mu\}|_{\lambda\rightarrow\infty}$ is the static permeability of the absorber. Other works have followed the approach presented in \cite{Rozanov2000} and have shown similar bounds \cite{Gustafsson2011, gustafsson2016physical} for: materials with open boundary conditions, \cite{xia2021theoretical} high impedance surfaces, \cite{Gustafsson2011} antenna arrays, \cite{Doane2013,jonsson2013array,doane2014bandwidth} and metamaterials \cite{sohl2008scattering,gustafsson2010sum,gustafsson2012optical}.

Here we derive a fundamental relationship between the absorption bandwidth and thickness for general transparent homogeneous layers. We consider the analytic properties of both the reflection and transmission coefficients which results in a dispersion relationship connecting the slab thickness and the static permeability and permittivity.

\section{Transfer Matrix for a Single Layer}
Consider a homogeneous slab of matter of thickness $d$ embedded in vacuum with material parameters $\epsilon=\epsilon_r \epsilon_0$ and $\mu=\mu_r \mu_0$, where $\epsilon_r$ and $\mu_r$ are the relative quantities, and $\epsilon_0$ and $\mu_0$ are the vacuum values. The reflection coefficient ($r$) and transmission coefficient ($t$) are determined using the transfer matrix method and at normal incidence they are,

\begin{equation}
    r=\frac{\displaystyle -\frac{i}{2}\left[Z_r^{-1}-Z_r \right]\sin \left(nk_0d\right)}{\displaystyle \cos \left(nk_0d\right)+\frac{i}{2}\left[Z_r^{-1}+Z_r \right]\sin \left(nk_0d \right)}
    \label{r_tmatrix}
\end{equation}

\noindent and

\begin{equation} \label{t_tmatrix}
    t=\frac{\displaystyle 1}{\displaystyle \cos \left(nk_0d\right)+\frac{i}{2}\left[Z_r^{-1}+Z_r \right]\sin \left(nk_0d \right)}
\end{equation}

\noindent where $n = \sqrt{\mu_r\epsilon_r}$ is the refractive index, $Z_r=\sqrt{\mu_r/\epsilon_r}$ is the relative impedance, and $k_0=2\pi/\lambda_0$ is the free-space wavevector.

Notice that $r$ and $t$ are related, i.e. they can be written, \cite{Markos2008} 

\begin{equation}
    \frac{r}{t}=-\frac{i}{2}\left[Z_r^{-1}-Z_r \right]\sin \left(nk_0d\right)
    \label{eq5}
\end{equation}

The magnitude squared of Eq. \ref{eq5} is therefore, 

\begin{equation}
    \left|\frac{r}{t}\right|^2=\frac{1}{4}\left[ \left| Z_r^{-1}-Z_r \right|^2 \right] \left| \sin{(nk_0d)} \right|^2
    \label{tr_mag}
\end{equation}

\noindent Equation \ref{tr_mag} is the main result from the transfer matrix that we use for $r$ and $t$ to bound the absorption.\\

\section{High frequency limit of material parameters}
We specify the values of material parameters -- where they exist -- at infinite frequency (zero wavelength). All material parameters at zero wavelength, or infinite frequency, must take on the values of free-space in the classical limit. Specifically,

\begin{equation}
\tilde{\epsilon}_r(\omega)|_{\omega\rightarrow\infty}=1 \quad \text{or} \quad \tilde{\epsilon}_r(\lambda)|_{\lambda\rightarrow 0}=1
\end{equation}

\begin{equation}
      \tilde{\mu}_r(\omega)|_{\omega\rightarrow\infty}=1 \quad \text{or} \quad \tilde{\mu}_r(\lambda)|_{\lambda\rightarrow 0}=1
\end{equation}

\begin{equation}
\tilde{n}(\omega)|_{\omega\rightarrow\infty}=1 \quad \text{or} \quad \tilde{n}(\lambda)|_{\lambda\rightarrow 0}=1
\end{equation}

\begin{equation}
\tilde{Z}_r(\omega)|_{\omega\rightarrow\infty}=1 \quad \text{or} \quad \tilde{Z}_r(\lambda)|_{\lambda\rightarrow 0}=1
\end{equation}

\noindent where the imaginary portion of each of these parameters is zero.

\section{Details of Calculations and the Dispersion Relation}
Calculations in Ref. \cite{Rozanov2000} are mainly focused on bounding $\int_{0}^{\infty} \ln (|r(\lambda)|) d \lambda$ since in that case $|r(\lambda)|^2  = 1 - A(\lambda)$, where $A(\lambda)$ is the absorption in wavelength $\lambda$. In our case $1 - A(\lambda) = |r(\lambda)|^2 + |t(\lambda)|^2$, where $t(\lambda)$ is the transmission coefficient at wavelength $\lambda$. In this light, we focus on bounding $\int_{0}^{\infty} \ln (|r(\lambda)|^2 + |t(\lambda)|^2) d \lambda$ instead.
Clearly

\begin{align*}
    \ln \left(|r(\lambda)|^2 + |t(\lambda)|^2\right) = \ln (|t(\lambda)|^2) +  \ln \left[1 + \left|\frac{r(\lambda)} {t(\lambda)}\right|^2\right].
\end{align*}

Thus invoking Eq. \ref{tr_mag} we find,

\begin{eqnarray*}
\int_{0}^{\infty} \ln \left(|r(\lambda)|^2 + |t(\lambda)|^2\right) d \lambda =
\int_{0}^{\infty} \ln (|t(\lambda)|^2) d \lambda + \\
+ \int_{0}^{\infty} \ln \left[1 + \frac{1}{4}\left[|Z_r^{-1}-Z_r|^2 \right] |\sin(nk_0d)|^2 \right] d \lambda
\end{eqnarray*}

\noindent We note that $|r(\lambda)|^2 + |t(\lambda)|^2 \le 1$ thus
$\int_{0}^{\infty} \ln (|t(\lambda)|^2) d \lambda \le 0$ and $\int_{0}^{\infty} \ln (|r(\lambda)|^2 + |t(\lambda)|^2) d \lambda \le 0$. Additionally,
\begin{align*}
    \int_{0}^{\infty} \ln (|t(\lambda)|^2) d \lambda \le \int_{0}^{\infty} \ln (|r(\lambda)|^2 + |t(\lambda)|^2) d \lambda \le 0
\end{align*}

\noindent assuming that (as we will prove later) these integrals exist.

\begin{figure}
    \centering
    \includegraphics[width=0.3\textwidth]{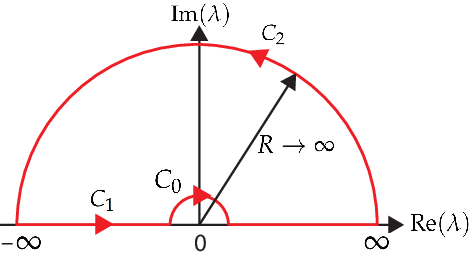}
    \caption{Contour used for integration of Eq. \ref{contour}}
    \label{Fig1}
\end{figure}

We next focus on $\int_{0}^{\infty} \ln (|t(\lambda)|^2) d \lambda$. The function \(t(\lambda)\) is causal and therefore analytic in the upper half-plane \cite{toll1956causality}. This analyticity implies that any poles present in the upper half-plane of \(\ln(t(\lambda))\) arise solely from the zeros of \(t(\lambda)\). As \(\lambda\) approaches 0, the slab's impedance approaches a value of \(Z_r = Z_0 = 1\), leading \(t\) to assume the form \(e^{i\infty}\). This infinite phase behavior at \(t(\lambda=0)\) introduces an intrinsic singularity for \(\ln(t(\lambda))\), which is distinct from other poles associated with zeros of \(t(\lambda)\). 

Therefore, to evaluate $\int_{0}^{\infty} \ln (|t(\lambda)|^2) d \lambda$ we perform a contour integral over the composite path $C = C_0 + C_1 + C_2$ as illustrated in Fig. \ref{Fig1}. As shown in the Supplemental, the contribution due to the intrinsic singularity (\(C_0 \) term) gives a residue of \(2\pi^2d\sqrt{\epsilon_\infty\mu_\infty}\), where $\epsilon_\infty \equiv \epsilon_{\lambda \rightarrow 0}$, $\mu_\infty \equiv \mu_{\lambda \rightarrow 0}$, and both $\epsilon_\infty$ and $\mu_\infty$ are real.

To address poles of $\ln(t(\lambda))$ which arise from the zeros of $t(\lambda)$, we assume \( n \) poles in the upper half-plane, denoted as \( \lambda_1, \lambda_2, \dots, \lambda_n \). Using a product expansion \cite{Nussenzveig_1972} we multiply $t(\lambda)$ with a Blaschke product and define an ancillary function as,

\begin{align*}
    t'(\lambda) \equiv t(\lambda) \prod^n_i\frac{(\lambda-\lambda^*_i)}{(\lambda-\lambda_i)} \equiv t(\lambda)B(\lambda)
\end{align*}

\noindent that has neither poles nor zeros in the upper half plane, where $\lambda^*$ denotes the complex conjugate of $\lambda$. $\ln(t'(\lambda))$ is analytic in the upper half plane of $\lambda$ since $t'(\lambda)$ does not have zeros in the upper half plane. Thus, from Cauchy's integral theorem, we have

\begin{equation}\label{contour}
    \oint_{\mathcal{C}} \ln(t'(\lambda)) \, d\lambda = 0
\end{equation}

Taking the real part of the integral along paths $C = C_0 + C_1 + C_2$, and noting that $\Re \{\ln [t'(\lambda)]\} = |\ln [t'(\lambda)]| = |\ln [t(\lambda)]|$ with the evenness of $|\ln [t(\lambda)]|$, we have

\begin{align*}
     2 \int_{0}^{\infty}  |\ln (t(\lambda))| d\lambda + \Re\oint_{\mathcal{C}_0}  \ln (t(\lambda)) d \lambda + \Re\oint_{\mathcal{C}_2}  \ln (t(\lambda)) d \lambda\\ + \Re\oint_{\mathcal{C}_0}  \ln (B(\lambda)) d \lambda + \Re\oint_{\mathcal{C}_2}  \ln (B(\lambda)) d \lambda = 0
\end{align*}

Building on the prior discussion and to simplify our analysis further, we will make the following assumptions: the limits of $\lim_{\lambda \rightarrow \infty} \epsilon_r(\lambda) \equiv \epsilon_s$  and $\lim_{\lambda \rightarrow \infty} \mu_r(\lambda) \equiv \mu_s$ exist.

Under the above assumption, as $|\lambda| \rightarrow \infty$, we may approximate Eq. S24 (see Supplemental) as,

\begin{equation}
   t(\lambda) \sim   1 - i  \frac{\pi d (\epsilon_s + \mu_s)}{\lambda} + O\left(\frac{1}{\lambda^2}\right) 
   \label{tapprox}
\end{equation}

\noindent where $O\left(\frac{1}{\lambda^2}\right)$ denotes terms of order $\frac{1}{\lambda^k}$ for $k \ge 2$ in the Laurent expansion of $t(\lambda)$ around the point at infinity in $P^+$.

We next calculate $\Re (\oint_{\mathcal{C}_2}  \ln (t(\lambda)) d \lambda )$ as $R \rightarrow \infty$. Letting $\lambda = R \exp(i \theta)$ where $\theta$ varies from $0$ to $\pi$, we have $d \lambda = i \lambda d \theta$. As $R$ grows large, using Eq. \ref{tapprox} we have
\begin{align*}
    \ln (t(\lambda)) \sim  \ln \left[1 - i  \frac{\pi d (\epsilon_s + \mu_s)}{\lambda} + O\left(\frac{1}{\lambda^2}\right)\right].
\end{align*}

\noindent Simple manipulation gives

\begin{align*}
    \lim_{R \to \infty} \Re \oint_{\mathcal{C}_2} \ln (t(\lambda)) d \lambda  =  \pi^2 d \Re(\epsilon_s + \mu_s)
\end{align*}

We use a similar approach to evaluate $\Re\oint_{\mathcal{C}_0}  \ln \left[t'(\lambda)\right] d \lambda$ and  $\Re\oint_{\mathcal{C}_2}  \ln \left[t'(\lambda)\right] d \lambda$ as $R \rightarrow 0$ and $R \rightarrow \infty$, respectively. Simple manipulation gives

\begin{align*}
    \lim_{R \to 0} \Re\oint_{\mathcal{C}_0}  \ln \left[B(\lambda)\right] d \lambda &= 0\\
    \lim_{R \to \infty} \Re\oint_{\mathcal{C}_2}  \ln \left[B(\lambda)\right] d \lambda &= -2\pi \sum_i\Im(\lambda_i)
\end{align*}

Plugging the contribution of each term back into Equation \ref{contour}, we have

\begin{align*}
    \int_{0}^{\infty}  \ln (|t(\lambda))|^2) d\lambda = &- \pi^2 (\epsilon_{s,r} + \mu_{s,r}) d \\ &+2\pi^2d\sqrt{\epsilon_\infty\mu_\infty} + 2\pi \sum_i\Im(\lambda_i)
\end{align*}

\noindent where $\epsilon_{s,r}\equiv\Re(\epsilon_s)$ and $\mu_{s,r}\equiv\Re(\mu_s)$. Since $\Im(\lambda_i) > 0$ in the upper half plane of $\lambda$

\begin{align*}
    \int_{0}^{\infty}  \ln (|t(\lambda))|^2) d\lambda \geq - \pi^2 (\epsilon_{s,r} + \mu_{s,r}) d +2\pi^2d\sqrt{\epsilon_\infty\mu_\infty}
\end{align*}

It follows that

\begin{multline*} -\pi^2 \bigl[\epsilon_{s,r} + \mu_{s,r}-2\sqrt{\epsilon_\infty\mu_\infty}\bigr]d \leq \int_{0}^{\infty} \ln (|t(\lambda)|^2) d \lambda \\
\le \int_{0}^{\infty} \ln (|r(\lambda)|^2 + |t(\lambda)|^2) d \lambda  \le 0
\end{multline*}

\noindent Which gives

\begin{multline*}
   \left|\int_{0}^{\infty} \ln (|r(\lambda)|^2 + |t(\lambda)|^2) d \lambda \right|\\
   \le \pi^2 \Bigl|\epsilon_{s,r}+\mu_{s,r}-2\sqrt{\epsilon_\infty\mu_\infty}\Bigr|d. 
\end{multline*}

\noindent or in terms of the material thickness -- redefine as $d_{TL}$,

\begin{equation}
    d_{TL} \ge \frac{\left|\int_{0}^{\infty} \ln \left[|r(\lambda)|^2 + |t(\lambda)|^2\right] d \lambda \right|}{\pi^2\Bigl|\epsilon_{s,r} + \mu_{s,r}-2\sqrt{\epsilon_\infty\mu_\infty}\Bigr|}\\
    \equiv d_{T}
    \label{result}
\end{equation}

This is an analog of the bound shown in \cite{Rozanov2000} and can be used in a corresponding manner to bound the absorption bandwidth as a function of thickness of transparent homogeneous materials. 

In particular if the absorption $A(\lambda)$  of the material between wavelengths  $[\lambda_0, \lambda_1]$ satisfies $A(\lambda) >  1 - \delta$ for some $0< \delta < 1$, then $|r(\lambda)|^2 + |t(\lambda)|^2 \le \delta$ in $[\lambda_0, \lambda_1]$ and it follows that
$|\ln(\delta)| (\lambda_1 - \lambda_0) \le  |\int_{0}^{\infty} \ln (|r(\lambda)|^2 + |t(\lambda)|^2) d \lambda | \le  \pi^2 |(\epsilon_{s,r} + \mu_{s,r})-2\sqrt{\epsilon_\infty\mu_\infty}| d$
and $d \ge  |\ln(\delta)| (\lambda_1 - \lambda_0) / [\pi^2 |(\epsilon_{s,r} + \mu_{s,r})-2\sqrt{\epsilon_\infty\mu_\infty}|]$.

\section{Validation}
We next validate Eq. \ref{result} for two test cases -- a homogeneous material of thickness $d$ modeled as a Lorentz oscillator, and numerical simulations of metamaterials.

\begin{figure*}[ptb]
    \centering
    \includegraphics[width=0.8\textwidth]{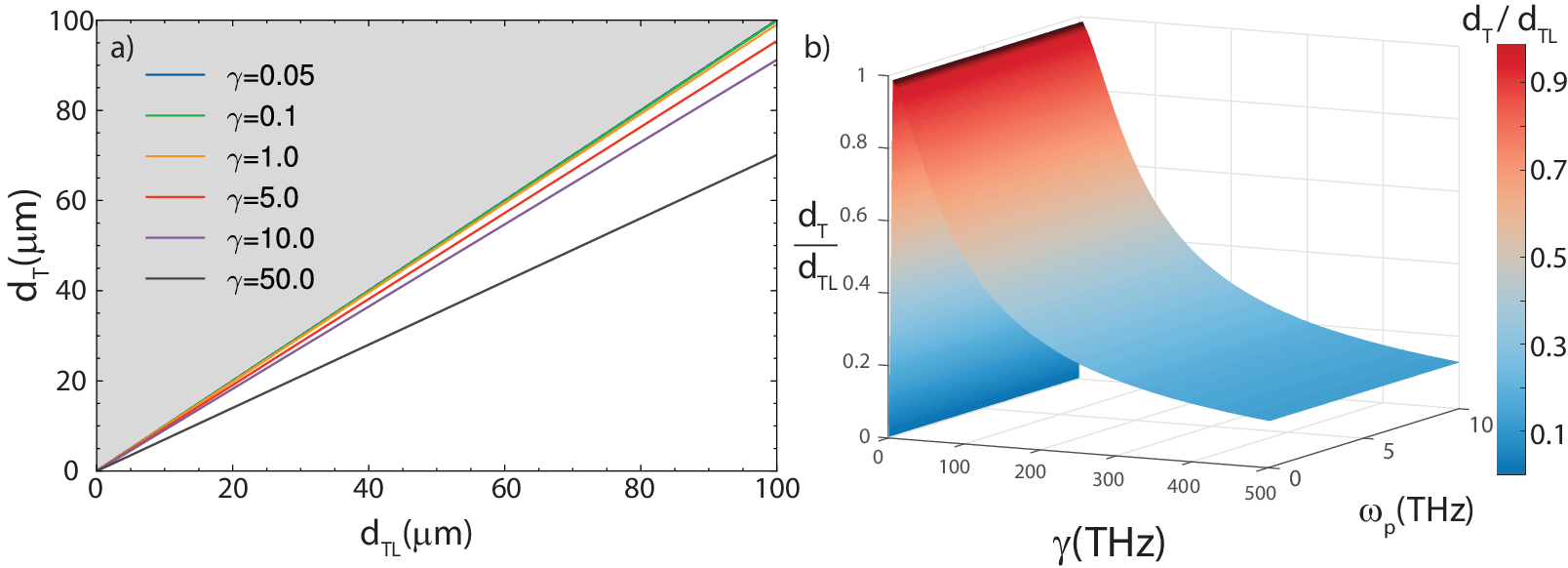}
    \caption{Calculated values of $d_{T}$ using a Lorentz model for a slab of magneto-dielectric material of thickness $d_{TL}$. In (a) the curves are for different damping frequencies $\gamma$, as described in the text. In (b), we plot the ratio $d_{T} / d_{TL}$ as a function of $\gamma$ and $\omega_p$ for the special case when $\tilde{\epsilon} = \tilde{\mu}$. }
    \label{fig2}
\end{figure*}

\subsection{Lorentz Oscillator Model}
The permittivity and permeability of the slab are described by the Lorentz oscillator and presented in the Supplemental.

We first assign equal permittivity and permeability values $\tilde{\epsilon} = \tilde{\mu}$, such that the slab impedance matches free space, i.e. $Z_r=1$. Therefore from Eqs. 3 and 4 we find,

\begin{align}
    r(\lambda)=&0 \label{tmmr} \\
    t(\lambda)=&\frac{1}{\cos{(nk_0d)}+i\sin{(nk_0d)}} \label{tmmt}
\end{align}

The slab thickness in the TMM is given as $d=d_{TL}$ in Eq. \ref{result}. Our objective is to verify the fundamental bound for transparent layers which asserts that $d_{TL} \geq d_{T}$. Here $d_{T}$ is calculated using the right side of Eq. \ref{result}, with $r(\lambda)$ and $t(\lambda)$ derived from Equations \ref{tmmr} and \ref{tmmt}, under the assumption that $\epsilon_s = \tilde{\epsilon}(\omega = 0)$ and $\mu_s = \tilde{\mu}(\omega = 0)$. The slab is given as: $\epsilon_\infty = 1.0$, $\omega_0 = 2\pi \times 1.0$ THz, $\omega_p = 2\pi \times 1.25$ THz, and $\gamma = 2\pi \times 0.05$ THz. In Fig. \ref{fig2} (a), we explore a range of damping frequencies (solid curves) from $\gamma = 2\pi \times 0.05$ THz to $\gamma = 2\pi \times 50$ THz, plotting $d_{LT}$ against $d_T$ while holding other parameters constant. None of the curves lie in the shaded gray area (where $d_{T} > d_{TL}$), which confirms the soundness of the specified limit. We further explored the dependence on the plasma frequency, a crucial Lorentz parameter for absorption, by varying $\omega_p$ from $2\pi \times 1.25$ THz to $2\pi \times 20.0$ THz. Notably, changes in $\omega_p$ did not influence the bound $d_T$, indicating that $\omega_p$ has no effect on the residual of $ln(t)$ in scenarios with an impedance-matched layer. The absence of any curves in the gray area reinforces the validity of Equation \ref{result}.

To further validate Eq.~\ref{result}, we examined the ratio $d_{T} / d_{TL}$ across a broad range of values for $\gamma$ and $\omega_p$, as depicted in Fig.~\ref{fig2}. Notably, for an impedance-matched slab, the real part of the residue terms solely depends on $\gamma$ and is unaffected by $\omega_p$. Across all the examined parameters, the value of $d_{T} / d_{TL}$ consistently remained below one, reinforcing the fundamental limit. Moreover, Fig.~\ref{fig2} showcases an inverse correlation between $d_{T}$ and $\gamma$. This observation suggests that a transmissive slab, when impedance-matched in free space, adheres to the anticipated absorptance behavior. As $\gamma$ decreases, the residual of $\ln(t)$ also diminishes, aligning the slab closer to the fundamental bound established by Eq.~\ref{result}.

\begin{figure}[ptb]
    \centering
    \includegraphics[width=1.0\columnwidth]{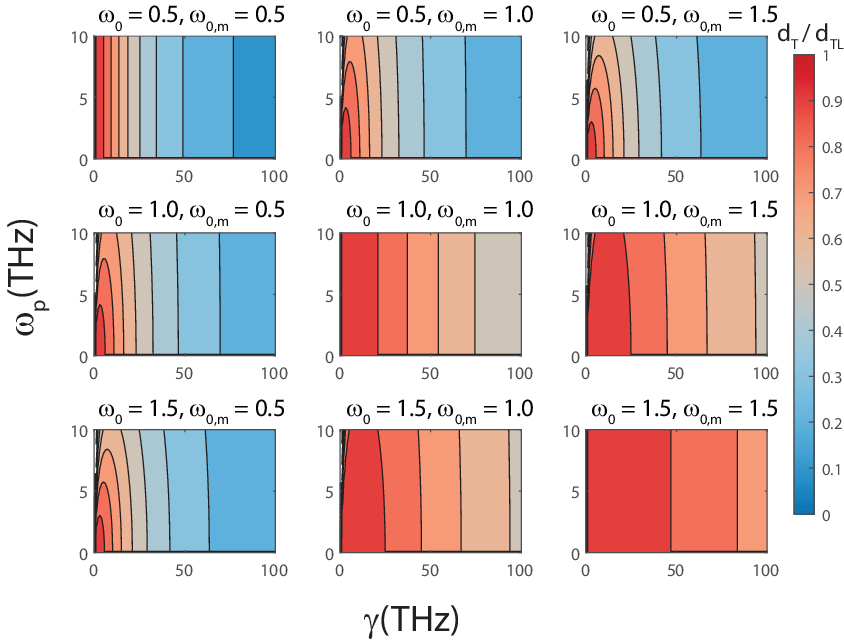}
    \caption{Contour plots of $d_T / d_{TL}$ for values of $\omega_0$ and $\omega_{0,m}$ from the Lorentz model, in which $\omega_p=\omega_{p,m}$ and $\gamma=\gamma_m$.}
    \label{fig3}
\end{figure}

While an impedance-matched transmissive slab provides substantial validation for Equation \ref{result}, it's important to note that our initial testing focused solely on the $\tilde{\epsilon} = \tilde{\mu}$ case. To broaden the scope of our verification process, we extended our test cases to include transmissive slabs with variable and non-equal $\tilde{\epsilon}$ and $\tilde{\mu}$ values. Fig. \ref{fig3} showcases contour plots of the $d_{T} / d_{TL}$ ratio with various values of $\gamma$ and $\omega_p$, where $\omega_p=\omega_{p,m}$ and $\gamma=\gamma_m$. Each of these plots represent a unique combination of resonance frequencies for $\tilde{\epsilon}$ and $\tilde{\mu}$. It becomes evident that altering the resonance frequency results in noticeable variations in $d_{T}$ across the domain. However, the bound as defined in Equation \ref{result}, specifically $d_{TL} \geq d_{T}$, remains unviolated.

\begin{figure}[ht]
    \centering
    \includegraphics[width=1.0\columnwidth]{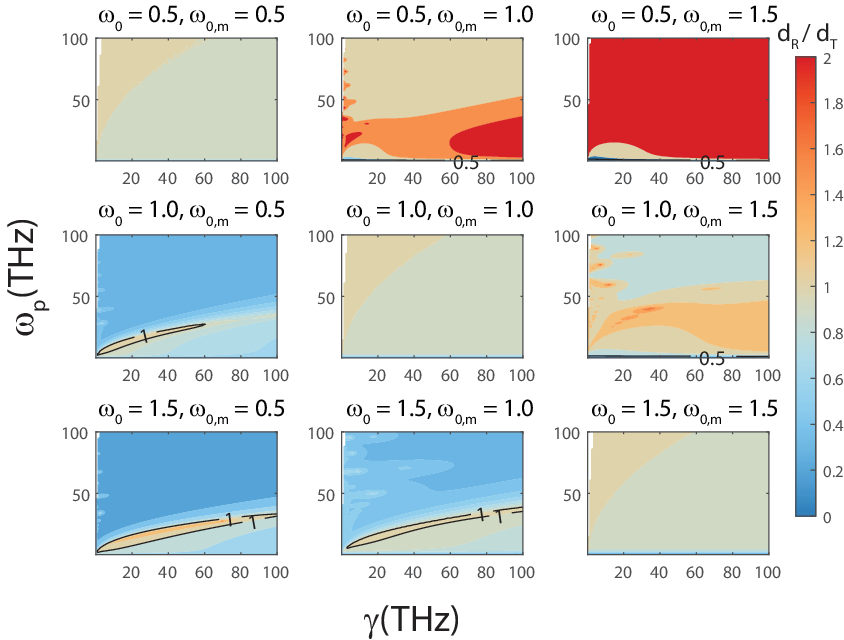}
    \caption{Contour plots of the ratio $d_{R} / d_{T}$ for values of $\omega_0$ and $\omega_{0,m}$ from the Lorentz model in which $\omega_p=\omega_{p,m}$ and $\gamma=\gamma_m$.}
    \label{fig4}
\end{figure}

Our research also investigates the relationship between the two fundamental limits discussed here, i.e. that of a transparent slab compared to that of a metal backed slab. The latter is defined on the right side of Eq. \ref{Rozanov_Lim} and denoted as $d_{R}$. Fig. \ref{fig4} displays contour plots that account for diverse combinations of $\gamma$, $\omega_p$, $\omega_0$, and $\omega_{0,m}$, with the contours and the color map given by the ratio $d_{R}/d_{T}$. The comparison of the differences between $d_{R}$ and $d_{T}$ is intriguing when the thickness is identical for both metal-backed and transmissive slabs, i.e. $d_{TL}=d_{RL}$. We generate contour plots for various configurations of $\gamma$, $\omega_p$, $\omega_0$, and $\omega_{0,m}$. Fig. \ref{fig4} shows color contour plots of $d_{R}/d_{T}$ versus $\omega_p$ and $\gamma$.

When both slabs are impedance-matched to free space, i.e. $\omega_0=\omega_{0,m}$, Fig. \ref{fig4} shows that $d_{R} = d_{T}$. This result may appear counterintuitive at first glance. Namely, when the two cases share identical material properties, absorption is solely governed by the optical path length within the slab. Considering that the metal-backed slab reflects the wave, the optical path length is double that of the transmissive slab. Yet, our findings elucidate this seemingly unconventional behavior. Our derivation of the fundamental limit diverges from the metal-backed limit approach by emphasizing reflectance $R$ and transmittance $T$ rather than the reflection coefficient $r$. In this context, the natural logarithm of the reflection coefficient is half the natural log of the reflectance. This aspect effectively counterbalances the doubled optical path length inherent in a metal-backed slab, resulting in a $d_{R}/d_{T}$ ratio that approaches unity. However, when $\omega_0 \ne \omega_{0,m}$, $d_{R}/d_{T}$ deviates from unity. For $\omega_0 < \omega_{0,m}$, $d_{R}/d_{T}$ may exceed 1 indicating a metal-backed slab achieves greater absorption bandwidth, while for $\omega_0 > \omega_{0,m}$, most $d_{R}/d_{T}$ remains below 1 where the transparent slab yields more absorption bandwidth. By elucidating these concepts, we aim to deepen the understanding of the relationship between Eqs. \ref{result} and \ref{Rozanov_Lim}, as well as the behaviors observed in Fig. \ref{fig4}.

\subsection{Electromagnetic Metamaterials}
Subwavelength metallic metamaterials play a pivotal role in absorbing electromagnetic radiation for myriad applications. \cite{padilla2022metamaterial,Liu2010b} Traditional configurations use a metallic split-ring resonator near a continuous ground plane, \cite{Schurig2006,Padilla2007b} which enhances performance by focusing light within a limited volume. Recent studies indicate that resonant modes supported by all-dielectric resonators also achieve subwavelength localization and high absorption. \cite{Cole2016,Liu2017} By exploiting the frequency degeneracy of these modes through careful metasurface design, highly efficient absorbers can be created. The use of metallic and dielectric materials exemplifies two fundamental strategies for constructing metasurface absorbers, and the absorption bandwidth serves as a critical metric for both systems. The limit established for a metal-back slab defines a core link between absorber thickness and absorption bandwidth. \cite{Rozanov2000} However, no universal theory currently exists outlining the relationship between absorption bandwidth and the absorber thickness for non metal-backed materials despite numerous theoretical, computational, and experimental studies. \cite{yu2019broadband,padilla2022metamaterial} Our results presented in Eq. \ref{result} introduces a fundamental thickness limit to the absorption bandwidth for metasurface absorbers, paving the way for more efficient design of transmissive absorbers.

We focused our analysis on metal-based and all-dielectric metamaterial (ADM) absorbers using Eq. \ref{result}. For both types of representative metamaterial systems, we chose commonly used resonator geometries for the periodic unit cell, tuned to give near-unity peak absorptivity at 1 THz. We also considered a 2$\times$2 supercell ADM, where two different unit cells can be combined to broaden the absorption bandwidth. We calculated $d_{T}$ by averaging the effective material parameters at low frequencies using known S-parameter inversion techniques. \cite{soukoulis2002} Details of the simulations and calculations can be found in the supplemental. Our computations are presented in Fig. \ref{fig5} for both ADM (\ref{fig5} (a)) and metal-based (\ref{fig5} (b)) metamaterial absorbers. Interestingly, all the metamaterials explored fall well short of the fundamental thickness to bandwidth limit, a result of their narrow-band absorption. Notably, the $2\times2$ resonator supercell, expected to broaden the ADM's absorptance peak, deviates further from the limit, suggesting higher values of $\mu_s$ and $\epsilon_s$.

\begin{figure}[ht]
    \centering
    \includegraphics[width=0.45\textwidth]{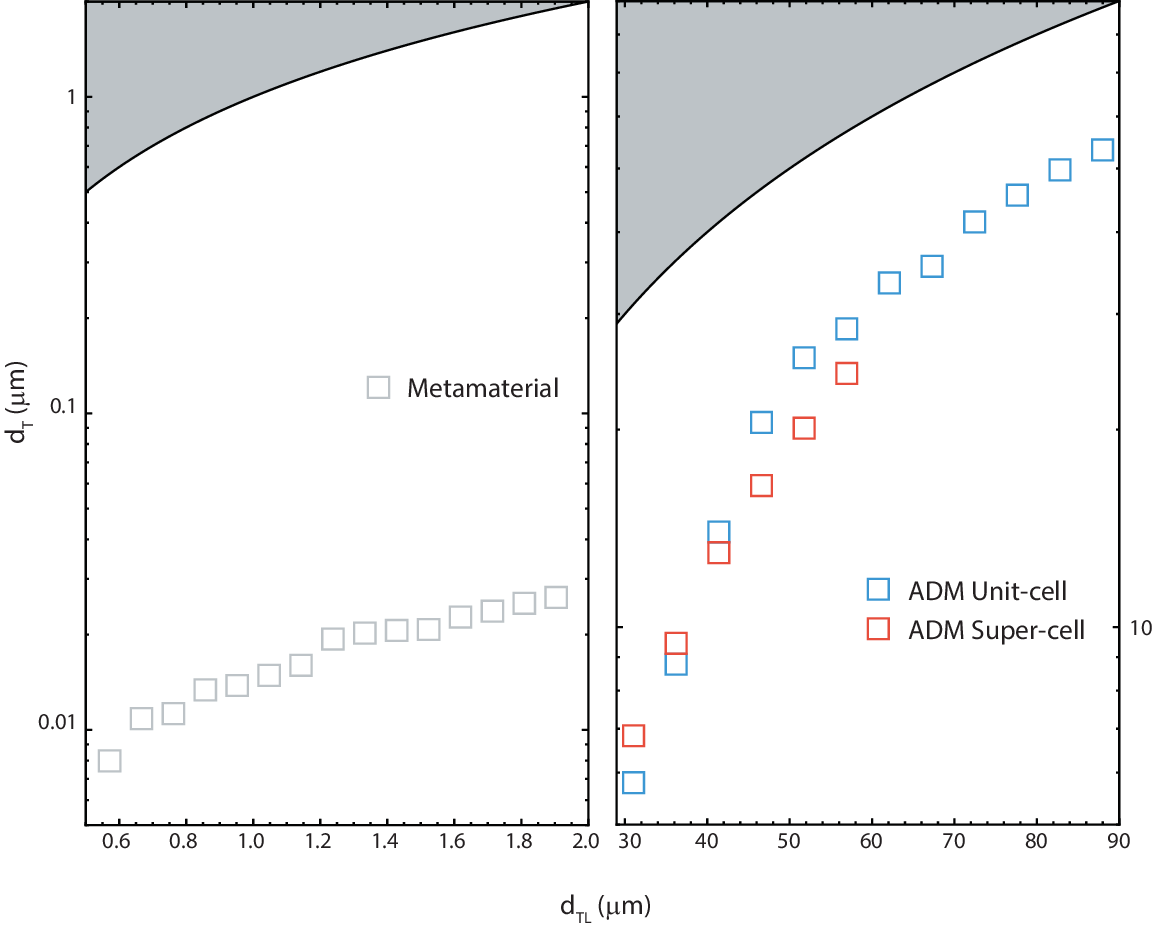}
    \caption{Calculated $d_{T}$ from metamaterial absorbers of thickness $d_{TL}$. The ADM unit cell is blue, the $2\times2$ ADM supercell is red. The metal-based unit-cell is shown as gray.}
    \label{fig5}
\end{figure}

\section{Conclusion}
Our investigation has extended fundamental relationships governing the absorption bandwidth and material thickness in non-metal backed layers. We have established a universal result based on the causal Kramers-Kronig relations, filling a significant gap in the current knowledge. This achievement was validated through transfer matrix calculations of homogeneous materials and full-wave numerical simulations of electromagnetic metallic and all-dielectric metamaterials, ensuring the robustness of our theory. Our results extend the known fundamental limits on absorbers, which is of crucial importance in both basic research and practical applications, and energy related fields stand to benefit greatly from this insight. We anticipate that this work will inform and inspire future studies in understanding the properties of absorptive materials and systems. We hope that our research can provide a robust theoretical foundation that can aid the design and development of more efficient energy harvesting and radiative systems, enabling more sustainable and cost-effective solutions to global energy needs.

\begin{acknowledgement}
We acknowledge useful discussions with Mats Gustafsson.
\end{acknowledgement}

\begin{funding}
We acknowledge support from the Department of Energy under U.S. Department of Energy (DOE) (DESC0014372).
\end{funding}

\begin{authorcontributions}
All authors have accepted responsibility for the entire content of this manuscript and approved its submission.

\end{authorcontributions}

\begin{conflictofinterest}
Authors state no conflict of interest.
\end{conflictofinterest}

\begin{dataavailabilitystatement}
The datasets generated during and/or analyzed during the current study are available from the corresponding author on reasonable request.
\end{dataavailabilitystatement}

\bibliographystyle{unsrt}
\bibliography{main}

\end{document}


\articletype{Research Article}

\author*[1]{Willie J. Padilla}
\author[2]{Yang Deng}
\author[3]{Omar Khatib}
\author[4]{Vahid Tarokh} 
\affil[1]{Pratt School of Engineering, Duke University, Durham, North Carolina,27705, USA, willie.padilla@duke.edu; https://orcid.org/0000-0001-7734-8847}
\affil[2]{Pratt School of Engineering, Duke University, Durham, North Carolina,27705, USA, omar.khatib@duke.edu; https://orcid.org/0000-0002-0721-9684}
\affil[3]{Pratt School of Engineering, Duke University, Durham, North Carolina,27705, USA, yang.deng@duke.edu; https://orcid.org/0000-0003-2249-6556}
\affil[4]{Pratt School of Engineering, Duke University, Durham, North Carolina,27705, USA, vahid.tarokh@duke.edu; ORCID}
\title{Supplemental Information for Fundamental absorption bandwidth to thickness limit for transparent homogeneous layers}
\runningtitle{Fundamental absorption bandwidth to thickness limit for transparent homogeneous layers}

\journalname{Nanophotonics}
\journalyear{2023}
\journalvolume{aop}

\maketitle

\section{Transfer Matrix Method}

\subsection{Single Layer Embedded in General Media}
We consider a flat slab of matter with a thickness $d$ which is sandwiched in-between two different infinite half spaces as depicted in Fig. S1. From left to right in Fig. S1 we denote these materials as Medium 1, Medium 2, and Medium 3, and each has material parameters of $(\mu_{r,1}, \epsilon_{r,1})$, $(\mu_{r,2}, \epsilon_{r,2})$, $(\mu_{r,3}, \epsilon_{r,3})$, respectively, or equivalently each media has a relative impedance given by $Z_{r,1}=\sqrt{\mu_{r,1}/\epsilon_{r,1}}$, $Z_{r,2}=\sqrt{\mu_{r,2}/\epsilon_{r,2}}$, and $Z_{r,3}=\sqrt{\mu_{r,3}/\epsilon_{r,3}}$. The transfer matrix for transverse electric (TE) polarization at normal incidence for the interface between Medium 1 and Medium 2 is given as,

\begin{figure}[ht]
    \centering
    \includegraphics[width=3.0in]{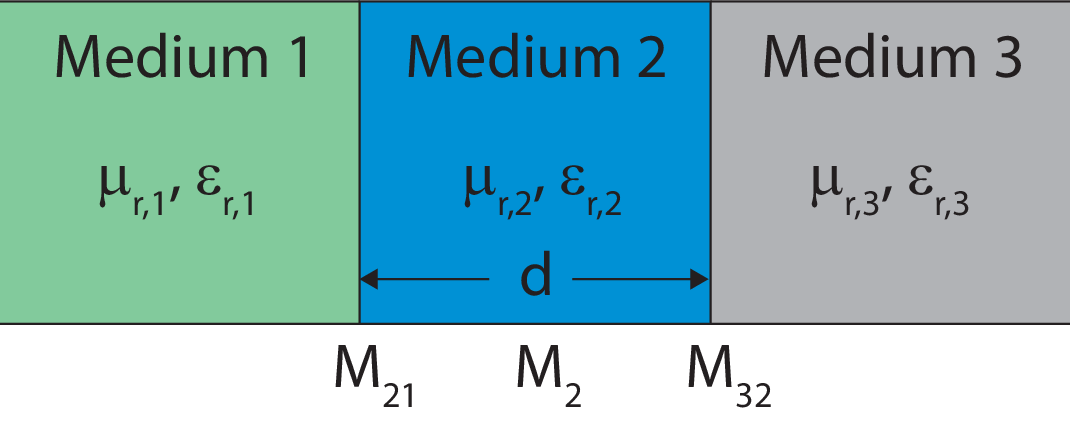}
    \caption{Illustration of a single flat slab sandwiched by two infinite half-spaces.}
    \label{fig:my_label}
\end{figure}

\begin{equation}
    M_{21}=\frac{1}{2}\begin{bmatrix}
      1+\frac{Z_{r,2}}{Z_{r,1}} & 1-\frac{Z_{r,2}}{Z_{r,1}} \\
      1-\frac{Z_{r,2}}{Z_{r,1}} & 1+\frac{Z_{r,2}}{Z_{r,1}} 
    \end{bmatrix}
\end{equation}  

\noindent And the TE transfer matrix between Medium 2 and Medium 3 is,

\begin{equation}
    M_{32}=\frac{1}{2}\begin{bmatrix}
      1+\frac{Z_{r,3}}{Z_{r,2}} & 1-\frac{Z_{r,3}}{Z_{r,2}} \\
      1-\frac{Z_{r,3}}{Z_{r,2}} & 1+\frac{Z_{r,3}}{Z_{r,2}} 
    \end{bmatrix}
\end{equation}

\noindent where $k=nk_0=2n\pi/\lambda_0=2\pi/\lambda$ is the wavevector inside the medium, $n$ is the refractive index, $k_0$ is the free-space wavevector, and $\lambda_0$ is the free-space wavelength. To align with the $e^{i\omega t}$ convention used in the main text, we define the transfer matrix for propagation across a thickness $d$ of Medium 2 is,

\begin{equation}
    M_{2}=\begin{bmatrix}
      e^{-ikd} & 0 \\
      0 & e^{ikd} 
    \end{bmatrix}
\end{equation}

\noindent The total transfer matrix is thus,

\begin{equation}
    M^{tot}=\begin{bmatrix}
      M^{tot}_{11} & M^{tot}_{12} \\
      M^{tot}_{21} & M^{tot}_{22} 
    \end{bmatrix}=M_{32} M_2 M_{21}
    \label{Mtot}
\end{equation}

\noindent where each component of $M^{tot}$ is,

\begin{align}
    M^{tot}_{11}=\frac{1}{2} \biggr [ \left(1+\frac{Z_{r,3}}{Z_{r,1}} \right) \cos{k_2d} -i \left(\frac{Z_{r,3}}{Z_{r,2}}+\frac{Z_{r,2}}{Z_{r,1}} \right) \sin{k_2d}\biggr ] \label{tm1} \\
    M^{tot}_{12}=\frac{1}{2} \biggr [ \left(1-\frac{Z_{r,3}}{Z_{r,1}} \right) \cos{k_2d} -i \left(\frac{Z_{r,3}}{Z_{r,2}}-\frac{Z_{r,2}}{Z_{r,1}} \right) \sin{k_2d}\biggr ] \label{tm2} \\
    M^{tot}_{21}=\frac{1}{2} \biggr [ \left(1-\frac{Z_{r,3}}{Z_{r,1}} \right) \cos{k_2d} +i \left(\frac{Z_{r,3}}{Z_{r,2}}-\frac{Z_{r,2}}{Z_{r,1}} \right) \sin{k_2d}\biggr ] \label{tm3} \\
    M^{tot}_{22}=\frac{1}{2} \biggr [ \left(1+\frac{Z_{r,3}}{Z_{r,1}} \right) \cos{k_2d} +i \left(\frac{Z_{r,3}}{Z_{r,2}}+\frac{Z_{r,2}}{Z_{r,1}} \right) \sin{k_2d}\biggr ] \label{tm4} \\
\end{align}

The reflection coefficient $r$ and transmission coefficient $t$ are given as,

\begin{equation}
    r=-\frac{M^{tot}_{21}}{M^{tot}_{22}}
    \label{rc}
\end{equation}

\noindent and

\begin{equation}
    t=\frac{detM^{tot}}{M^{tot}_{22}}
    \label{tc}
\end{equation}

For the specific case studied these are,

\begin{equation}
    r=\frac{-\left(1-\frac{Z_{r,3}}{Z_{r,1}}\right)\cos{k_2d_2}-i\left( \frac{Z_{r,3}}{Z_{r,2}}-\frac{Z_{r,2}}{Z_{r,1}} \right) \sin{k_2d_2} }{\left(1+\frac{Z_{r,3}}{Z_{r,1}}\right)\cos{k_2d_2}+i\left( \frac{Z_{r,3}}{Z_{r,2}}+\frac{Z_{r,2}}{Z_{r,1}} \right) \sin{k_2d_2} }
\end{equation}

\noindent and

\begin{equation}
    t=\frac{2 \frac{Z_{r,3}}{Z_{r,1}}}{\left(1+\frac{Z_{r,3}}{Z_{r,1}}\right)\cos{k_2d_2}+i\left( \frac{Z_{r,3}}{Z_{r,2}}+\frac{Z_{r,2}}{Z_{r,1}} \right) \sin{k_2d_2} }
\end{equation}

\subsection{Metal-Backed Single Layer}
A general expression for a layer of material described by permittivity $\epsilon_r=\epsilon/\epsilon_0$ and permeability $\mu_r=\mu/\mu_0$ of thickness $d$, lying on top of a conductor, may be obtained using the transfer matrix Eq. \ref{Mtot}. Here we take Medium 1 to be free-space, $\mu_{r,1}=\mu_0$ and $\epsilon_{r,1}=\epsilon_0$ -- or equivalently $Z_{r,1}=1$ -- and we take Medium 3 to be a perfect electric conductor with $Z_{r,3}=0$. Under these assumptions the components of the total transfer matrix of Eq. \ref{Mtot} becomes, 

\begin{align*}
    M_{11}^{tot}&=\frac{1}{2} \left(\cos{kd}-iZ_r\sin{kd}\right) \\
    M_{12}^{tot}&=\frac{1}{2} \left(\cos{kd}+iZ_r\sin{kd} \right) \\
    M_{21}^{tot}&=\frac{1}{2} \left(\cos{kd}-iZ_r\sin{kd} \right)\\
    M_{22}^{tot}&=\frac{1}{2}\left(\cos{kd}+iZ_r\sin{kd} \right) \\
\end{align*}

\noindent where we have dropped the sub-scripts since we only have one medium of thickness $d$ (formerly Medium 2).

The relative impedance and refractive index are respectively given by,

\begin{equation}
    Z_r=\sqrt{\mu_r / \epsilon_r}
    \label{impedance}
\end{equation}
\begin{equation}
    n=\sqrt{\epsilon_r \mu_r}.
    \label{index}
\end{equation}

The reflectance coefficient Eq. \ref{rc} and transmission coefficient Eq. \ref{tc} are,

\begin{equation}
    r=\frac{iZ_r\sin{kd}- \cos{kd}}{iZ_r\sin{kd}+ \cos{kd}},
\end{equation}

\noindent and

\begin{equation}
    t=0,
\end{equation}

\noindent In the longwavelength limit ($kd<<1$), the reflection coefficient becomes,
\begin{equation}
     r=-1+i\frac{4\pi\mu d}{\lambda_0}    
\end{equation}

\noindent and thus we recover Eq. 1 in main text.

\subsection{Single Layer Embedded in Vacuum}
If we next consider that the slab of material (Medium 2) is embedded in vacuum with $\mu_{r,1}=\mu_{r31}=\mu_0$ and $\epsilon_{r,1}=\epsilon_{r,3}=\epsilon_0$ -- thus $Z_{r,1}=Z_{r,3}=1$ -- we find elements of the total transmission matrix $M^{tot}$ of,

\begin{align}
    M^{tot}_{11}=&\frac{1}{2} \biggr [ 2\cos{kd} -i \left(\frac{1}{Z_{r}}+Z_{r} \right) \sin{kd}\biggr ] \\
    M^{tot}_{12}=&\frac{1}{2} \biggr [ -i \left(\frac{1}{Z_{r}}-Z_{r} \right) \sin{kd}\biggr ] \\
    M^{tot}_{21}=&\frac{1}{2} \biggr [ i \left(\frac{1}{Z_{r}}-Z_{r} \right) \sin{kd}\biggr ] \\
    M^{tot}_{22}=&\frac{1}{2} \biggr [ 2\cos{kd} +i \left(\frac{1}{Z_{r}}+Z_{r} \right) \sin{kd}\biggr ]
\end{align}

\noindent where again we have dropped the sub-scripts since there is only one layer (Medium 2) embedded in free-space. The reflection coefficient and transmission coefficient are,

\begin{equation}
    r=\frac{\displaystyle -\frac{i}{2}\left[Z_r^{-1}-Z_r \right]\sin \left(nk_0d\right)}{\displaystyle \cos \left(nk_0d\right)+\frac{i}{2}\left[Z_r^{-1}+Z_r \right]\sin \left(nk_0d \right)}
    \label{r_tmatrix}
\end{equation}

\noindent and

\begin{equation} \label{t_tmatrix}
    t=\frac{\displaystyle 1}{\displaystyle \cos \left(nk_0d\right)+\frac{i}{2}\left[Z_r^{-1}+Z_r \right]\sin \left(nk_0d \right)}
\end{equation}

\noindent Equations \ref{r_tmatrix} and \ref{t_tmatrix} may be used to solve for the index of refraction and relative impedance,

\begin{equation}
    n=\frac{1}{k_0d}\arcsin \frac{1}{2\tilde{t}}\left[1-\tilde{r}^2+\tilde{t}^2 \right]
\end{equation}

\begin{equation}
    Z_r=\pm\sqrt{\frac{(1+\tilde{r})^2-\tilde{t}^2}{(1-\tilde{r})^2-\tilde{t}^2}}
\end{equation}

\noindent The relative permeability and relative permittivity can then be determined from,

\begin{equation}
    \mu_r=nZ_r
\end{equation}

\noindent and

\begin{equation}
    \epsilon_r=\frac{n}{Z_r}
\end{equation}

Notice that $r$ and $t$ are related, i.e. they can be written,

\begin{equation}
    \frac{r}{t}=-\frac{i}{2}\left[Z_r^{-1}-Z_r \right]\sin \left(nk_0d\right)
\end{equation}

In the longwavelength limit ($kd<<1$), we may approximate,

\begin{equation}
    \frac{r}{t}=-\frac{i}{2}\left[Z_r^{-1}-Z_r \right]nk_0d
\end{equation}

Separately we find that $r$ in the long wavelength limit is,

\begin{equation}
    r=-\frac{ik_0d}{2}\left(\epsilon_r-\mu_r \right) = -\frac{i\pi d}{\lambda_0}\left(\epsilon_r-\mu_r \right)
\end{equation}

and $t$ in the long wavelength limit is,

\begin{equation}
    t=1-\frac{ik_0d}{2}\left(\epsilon_r+\mu_r \right) = 1-\frac{i\pi d}{\lambda_0}\left(\epsilon_r+\mu_r \right)
\end{equation}

\subsection{Two Layers Embedded in General Media}
We approach the two-layer problem by first considering the transfer matrix at the interface. Considering a single interface that has two materials on each side with material parameters $\epsilon_1, \mu_1$ and $\epsilon_2, \mu_2$. Then, we define the transfer matrix as:

\begin{align*}
    M = \frac{1}{2}\begin{pmatrix}
1+\frac{Z_{r,2}}{Z_{r,1}} & 1-\frac{Z_{r,2}}{Z_{r,1}}\\
1-\frac{Z_{r,2}}{Z_{r,1}} & 1+\frac{Z_{r,2}}{Z_{r,1}}
\end{pmatrix}
\end{align*}

where $Z_r$ is the relative impedance defined as $Z_r = \frac{Z}{Z_0}$. From a single interface, we can formulate the two-layers slabs embedded in vacuum as three interfaces problem:

\begin{align*}
    M^{tot} = M_{34}M_{3}M_{23}M_{2}M_{12}
\end{align*}

where $M_{34}, M_{23}, M_{12}$ represent interface between Medium 3\&4, 2\&3, 1\&2, respectively. $M_{3}$ and $M_{2}$ are the propagation matrix in Medium 3 and 2. Following the transfer matrix for interface, we expand $M^{tot}$ to:

\begin{align*}
    M^{tot} = \frac{1}{8}\begin{pmatrix}
1+\frac{Z_{r,4}}{Z_{r,3}} & 1-\frac{Z_{r,4}}{Z_{r,3}}\\
1-\frac{Z_{r,4}}{Z_{r,3}} & 1+\frac{Z_{r,4}}{Z_{r,3}}
\end{pmatrix}\begin{pmatrix}
e^{ik_3d_3} & 0\\
0 & e^{-ik_3d_3}
\end{pmatrix}\begin{pmatrix}
1+\frac{Z_{r,3}}{Z_{r,2}} & 1-\frac{Z_{r,3}}{Z_{r,2}}\\
1-\frac{Z_{r,3}}{Z_{r,2}} & 1+\frac{Z_{r,3}}{Z_{r,2}}
\end{pmatrix}\begin{pmatrix}
e^{ik_2d_2} & 0\\
0 & e^{-ik_2d_2}
\end{pmatrix} \\
\times\begin{pmatrix}
1+\frac{Z_{r,2}}{Z_{r,1}} & 1-\frac{Z_{r,2}}{Z_{r,1}}\\
1-\frac{Z_{r,2}}{Z_{r,1}} & 1+\frac{Z_{r,2}}{Z_{r,1}}
\end{pmatrix}
\end{align*}

\noindent where $Z_{r,i}$, $k_i$, and $d_i$ are the relative impedance, wavenumber, and slab thickness for $i^{th}$ medium. We write $M^{tot}$ as:

\begin{align*}
    M^{tot} = \begin{pmatrix}
M_{11} & M_{12}\\
M_{21} & M_{22}
\end{pmatrix}
\end{align*}

where,

\begin{align*}
    M_{11}^{tot} =& \frac{1}{2} \biggr [  \left(1+\frac{Z_{r,4}}{Z_{r,1}}\right)\cos{(k_3d_3)}\cos{(k_2d_2)}+i\left(\frac{Z_{r,4}}{Z_{r,3}}+\frac{Z_{r,3}}{Z_{r,1}}\right)\sin{(k_3d_3)}\cos{(k_2d_2)}\\
    &+i\left(\frac{Z_{r,4}}{Z_{r,2}}+\frac{Z_{r,2}}{Z_{r,1}}\right)\cos{(k_3d_3)}\sin{(k_2d_2)}-\left(\frac{Z_{r,3}}{Z_{r,2}}+\frac{Z_{r,4}Z_{r,2}}{Z_{r,3}Z_{r,1}}\right)\sin{(k_3d_3)}\cos{(k_2d_2)} \biggr ]
\end{align*}

\begin{align*}
    M_{12}^{tot} =& \frac{1}{2} \biggr [ \left(1-\frac{Z_{r,4}}{Z_{r,1}} \right)\cos{(k_3d_3)}\cos{(k_2d_2)}+i \left(\frac{Z_{r,4}}{Z_{r,3}}-\frac{Z_{r,3}}{Z_{r,1}}\right) \sin{(k_3d_3)}\cos{(k_2d_2)}\\
    &+i\left(\frac{Z_{r,4}}{Z_{r,2}}-\frac{Z_{r,2}}{Z_{r,1}}\right)\cos{(k_3d_3)}\sin{(k_2d_2)}-\left(\frac{Z_{r,3}}{Z_{r,2}}-\frac{Z_{r,4}Z_{r,2}}{Z_{r,3}Z_{r,1}}\right)\sin{(k_3d_3)}\cos{(k_2d_2)} \biggr]
\end{align*}

\begin{align*}
    M_{21}^{tot} =& \frac{1}{2}\biggr[\left(1-\frac{Z_{r,4}}{Z_{r,1}}\right)\cos{(k_3d_3)}\cos{(k_2d_2)}-i\left(\frac{Z_{r,4}}{Z_{r,3}}-\frac{Z_{r,3}}{Z_{r,1}}\right)\sin{(k_3d_3)}\cos{(k_2d_2)}\\
    &-i\left(\frac{Z_{r,4}}{Z_{r,2}}-\frac{Z_{r,2}}{Z_{r,1}}\right)\cos{(k_3d_3)}\sin{(k_2d_2)}-\left(\frac{Z_{r,3}}{Z_{r,2}}-\frac{Z_{r,4}Z_{r,2}}{Z_{r,3}Z_{r,1}}\right)\sin{(k_3d_3)}\cos{(k_2d_2)}\biggr]
\end{align*}

\begin{align*}
    M_{22}^{tot} =& \frac{1}{2}\biggr[\left(1+\frac{Z_{r,4}}{Z_{r,1}}\right)\cos{(k_3d_3)}\cos{(k_2d_2)}-i\left(\frac{Z_{r,4}}{Z_{r,3}}+\frac{Z_{r,3}}{Z_{r,1}}\right)\sin{(k_3d_3)}\cos{(k_2d_2)}\\
    &-i\left(\frac{Z_{r,4}}{Z_{r,2}}+\frac{Z_{r,2}}{Z_{r,1}}\right)\cos{(k_3d_3)}\sin{(k_2d_2)}-\left(\frac{Z_{r,3}}{Z_{r,2}}+\frac{Z_{r,4}Z_{r,2}}{Z_{r,3}Z_{r,1}}\right)\sin{(k_3d_3)}\cos{(k_2d_2)}\biggr]
\end{align*}

The reflection and transmission coefficients are then given by,

\begin{equation}
    r=-\frac{M_{21}^{tot}}{M_{22}^{tot}}
\end{equation}

\noindent and

\begin{equation}
    t=\frac{1}{M_{22}^{tot}}
\end{equation}

\section{Lorentz Oscillator Model}

We consider a general magneto-dielectric slab where the constitutive material parameters are represented by a Lorenz oscillator model:

\begin{align*}
\tilde{\epsilon}&=\epsilon_\infty+\frac{\omega_p^2}{\omega_0^2-\omega^2-i\gamma\omega}\\
\tilde{\mu}&=\mu_\infty+\frac{\omega_{p,m}^2}{\omega_{0,m}^2-\omega^2-i\gamma_m\omega}\\
\end{align*}

\noindent where $\epsilon_\infty, \mu_\infty$ are the permittivity and permeability, respectively, at infinite frequency, $\omega_p, \omega_{p,m}$ are the plasma frequencies, $\omega_0, \omega_{0,m}$ are the oscillator's center frequencies, and$\gamma, \gamma_m$ are the damping frequencies. We use a sub-scripted 'm' to distinguish the Lorentz parameters for the $\mu$ oscillator.


\section{High and low frequency limit of material parameters}
We specify the values of material parameters -- where they exist -- at both zero frequency (infinite wavelength) and infinite frequency (zero wavelength). All material parameters at zero wavelength or infinite frequency must take on the values of free-space in the classical limit. Specifically,

\begin{equation}
     \tilde{\epsilon}_r(\omega)|_{\omega\rightarrow\infty}=1 \quad \text{or} \quad \tilde{\epsilon}_r(\lambda)|_{\lambda\rightarrow 0}=1
\end{equation}
\begin{equation}
     \tilde{\mu}_r(\omega)|_{\omega\rightarrow\infty}=1 \quad \text{or} \quad \tilde{\mu}_r(\lambda)|_{\lambda\rightarrow 0}=1
\end{equation}
\begin{equation}
     \tilde{n}(\omega)|_{\omega\rightarrow\infty}=1 \quad \text{or} \quad \tilde{n}(\lambda)|_{\lambda\rightarrow 0}=1
\end{equation}
\begin{equation}
     \tilde{Z}_r(\omega)|_{\omega\rightarrow\infty}=1 \quad \text{or} \quad \tilde{Z}_r(\lambda)|_{\lambda\rightarrow 0}=1
\end{equation}

\noindent where the imaginary portion of each of these parameters is zero.

At the other end of the spectrum, i.e. at zero frequency or as $\lambda \rightarrow \infty$, it is not possible to specify any bounds on the material parameters. For example, at $\omega=0$ the real part of the relative impedance of a conductor is $\text{Re}(Z_r)=0$, but for a non-magnetic dielectric it may have a value of $\text{Re}(Z_r)=0.5$, and for a strong magnetic material $\text{Re}(Z_r)=300$.

\section{Material Parameters and Dispersion Bounds}
Here we define the material parameters -- sometimes termed optical constants -- and show the relationships between them. All parameters are complex and here we use notation where sub-scripted '1' indicates the real part, and a '2' the complex part. All parameters are also function of angular frequency $\omega=2\pi f$ where $\lambda f=c$ and $c$ is the speed of light in vacuum. Thus the permittivity ($\tilde{\epsilon}$) and permeability ($\tilde{\mu}$) are,

\begin{equation}
    \tilde{\epsilon}(\omega)=\epsilon_1(\omega)+i \epsilon_2(\omega)
\end{equation}

\begin{equation}
    \tilde{\mu}(\omega)=\mu_1(\omega)+i \mu_2(\omega)
\end{equation}

The material parameters may be expressed in relative unit-less quantities ($\epsilon_r,\mu_r$), defined through,

\begin{equation}
    \tilde{\epsilon}(\omega)=\epsilon_0 \tilde{\epsilon}_r(\omega)
\end{equation}

\begin{equation}
    \tilde{\mu}(\omega)=\mu_0 \tilde{\mu}_r(\omega)
\end{equation}
 
\noindent where $\epsilon_0=8.85 \times 10^{-12}$ F/m, and $\mu_0=4\pi \times 10^{-7}$ H/m are the values (SI units) in the vacuum, and the sub-scripted '$r$' indicates the relative quantity. 

The only two quantities that appear directly in Maxwell's equations are $\epsilon$ and $\mu$, and all other materials parameters have simple algebraic relationships to them. The wave impedance is,

\begin{equation}
    \tilde{Z}(\omega)=Z_1+i Z_2=\sqrt{\frac{\tilde{\mu}}{\tilde{\epsilon}}}
\end{equation}

\noindent We may also define a relative wave impedance as,

\begin{equation}
    \tilde{Z}(\omega)=Z_0 \tilde{Z}_r(\omega)
\end{equation}

\noindent where $Z_0=377$ Ohms is the value of free-space. 

The index of refraction is defined through the relative permittivity and relative permeability by,

\begin{equation}
    \tilde{n}(\omega)=n_1(\omega)+in_2(\omega)=\sqrt{\tilde{\epsilon}_r(\omega)\tilde{\mu}_r(\omega)}
\end{equation}

The wavevector $k$ is given by,

\begin{equation}
    \tilde{k}(\omega)=k_1(\omega)+ik_2(\omega)=\frac{\omega}{c}n(\omega)=k_0n(\omega)
\end{equation}
 
\noindent where we have defined the free-space wavevector $k_0=\omega / c$. The real part of the wavevector is related to the free-space wavelength through,

\begin{equation}
    k_1=\frac{2\pi}{\lambda_0}
\end{equation}

\noindent where we have explicitly noted the free-space wavelength $\lambda_0$. The wavelength in matter is,

\begin{equation}
    \lambda=\frac{\lambda_0}{n}
\end{equation}
 
We only consider matter which may be described as $\mathbf{B}=\mu \mathbf{H}$ and $\mathbf{D}=\epsilon \mathbf{E}$. Thus the energy density from Poynting's theorem may be written in the frequency domain as,

\begin{equation}
    U = \frac{\partial (\epsilon \omega)}{\partial \omega}\mathbf{E}\cdot \mathbf{E}^* + \frac{\partial (\mu \omega)}{\partial \omega}\mathbf{H}\cdot \mathbf{H}^* \geq 0
\end{equation}

\noindent Both $\epsilon$ and $\mu$ must therefore be functions of frequency, otherwise the energy can be negative when both $\epsilon$ and $\mu$ are negative. Therefore the permittivity and permeability must satisfy,

\begin{equation}
    \frac{\partial \omega\epsilon(\omega) }{\partial \omega} >0
\end{equation}

\begin{equation}
    \frac{\partial \omega\mu(\omega) }{\partial \omega} >0
\end{equation}


\section{Comment on Calculation of Dispersion Relation}

Here we note that in the derivation of the new thickness bound for a transparent homogeneous layer in the main text, the value of $\int_{0}^{\infty} \ln [1 + |r(\lambda) / t(\lambda)|^2] d \lambda$ was lower bounded by zero.  A natural question that arises is that if the consideration of this term can improve the above bound on ultimate thickness?  We note that when $\lambda \in [\lambda_1 , \lambda_2]$ for large values of $0 << \lambda_1 < \lambda_2$, it is easy to see that
\begin{align*}
    \int_{\lambda_1}^{\lambda_2} \ln (1 + \frac{1}{4}\left[|Z_r^{-1}-Z_r|^2 \right] |\sin(nk_0d)|^2 ) d \lambda = O(\frac{1}{\lambda_1})
\end{align*}

and can be made arbitrarily small for large values of $\lambda_1$. This means that the contribution of this term to the bound in large bandwidth regimes is extremely small and can thus be ignored.

\section{Treatment of Singularity for $t(\lambda = 0)$}
We revisit the transmission coefficient $t(\lambda)$ derived from the transfer matrix for a slab of thickness $d$, given by,

\begin{equation}\label{t_tm}
        t=\frac{\displaystyle 1}{\displaystyle \cos \left(kd\right)-\frac{i}{2}\left[Z_r^{-1}+Z_r \right]\sin \left(kd \right)}
\end{equation}

\noindent Here $k=k_0n$ where $n$ is the refractive index, and $k_0$ is the free-space wavevector. The transmission coefficient $t(\lambda)$ is analytic in the upper half plane and from Eq. \ref{t_tm} it can be shown that $t(\lambda)$ has no zeros in the upper half plane.

However, $t(\lambda)$ still possesses a point in the upper half plane which is not well-defined. It is instructive to write $t(\lambda)$ as \cite{heavens1991optical},

\begin{equation}\label{t}
        t=\frac{e^{-ikd}t_1t_2}{1+e^{-i2kd}r_1r_2}
\end{equation}

where,

\begin{align*}
    t_1 = \frac{2z}{z+1}, t_2 = \frac{2}{z+1}\\
    r_1 = \frac{z-1}{z+1}, r_2 = \frac{1-z}{z+1}\\
\end{align*}

\noindent Here $t_1$ ($r_1$) and $t_2$ ($r_2$) are the transmission coefficients (reflection coefficients) for the first and second interfaces, respectively, of the slab. 

At $\lambda = 0$, $e^{ikd}=e^{i\infty}$ which is undefined. To accommodate for the undefined functional behavior at $t(\lambda =0)$, we consider a contour integral. Let C$_0$ denote the semicircular path in Fig. \ref{Fig1}, that is an infinitesimally small semicircle around the origin of the complex plane. As required by causality, the permittivity $\epsilon$ and permeability $\mu$ values must approach the vacuum values for $\lambda \rightarrow \infty$. For the relative permittivity and permeability causality thus requires,

\begin{align*}
    \epsilon_{\lambda\rightarrow0} = 1, \mu_{\lambda\rightarrow0} = 1
\end{align*}

Using these assumptions, Eq. \ref{t} simplifies to:

\begin{align*}
t_{\lambda\rightarrow0} = e^{-ikd}.
\end{align*}

Letting $\lambda=\rho \exp{(i\theta)}$ we find for the integral of ln$(t(\lambda))$,

\begin{align*}
    \oint_{\mathcal{C}_0} \ln \left[t(\lambda)\right] d \lambda &= \int_\pi^0  \ln \left[ t(\rho e^{i\theta})\right]i\rho e^{i\theta}d\theta\\
    &= \int_\pi^0  \ln \left[ \text{exp} \left( -i\frac{2\pi nd}{\rho e^{i\theta}} \right)  \right] i\rho e^{i\theta} d\theta\\
    &= -2\pi^2d\sqrt{\epsilon_{\lambda\rightarrow0}\mu_{\lambda\rightarrow0}}\\
\end{align*}

\section{Evaluation of integrals on $C_0$ and $C_2$}

We expand the simple manipulations in the main text as follows:

\begin{align*}
    \Re\oint_{\mathcal{C}_0}  \ln \left[\prod^n_i\frac{(\lambda-\lambda^*_i)}{(\lambda-\lambda_i)}\right] d \lambda &= \Re\oint_{\mathcal{C}_0}  \sum^n_i\ln \left[1+\frac{2\Im(\lambda_i)}{(\lambda-\lambda_i)}\right] d \lambda\\
    &= \lim_{R \to 0}\Re\int^{\pi}_{0}  \sum^n_i\ln \left[1+\frac{2\Im(\lambda_i)}{(Re^{i\theta}-\lambda_i)}\right] iRe^{i\theta} d \theta\\
    &= 0 
\end{align*}

\begin{align*}
    \Re\oint_{\mathcal{C}_2}  \ln \left[\prod^n_i\frac{(\lambda-\lambda^*_i)}{(\lambda-\lambda_i)}\right] d \lambda &= \Re\oint_{\mathcal{C}_2}  \sum^n_i\ln \left[1+\frac{i2\Im(\lambda_i)}{(\lambda-\lambda_i)}\right] d \lambda\\
    &= \lim_{R \to \infty}\Re\int^{\pi}_{0}  \sum^n_i\ln \left[1+\frac{i2\Im(\lambda_i)}{(Re^{i\theta}-\lambda_i)}\right] iRe^{i\theta} d \theta\\
    &= \pi \sum^n_i-2\Im(\lambda_i)
\end{align*}

\section{Transfer Matrix Method Validation}

\begin{figure}[ht]
    \centering
    \includegraphics[width=\textwidth]{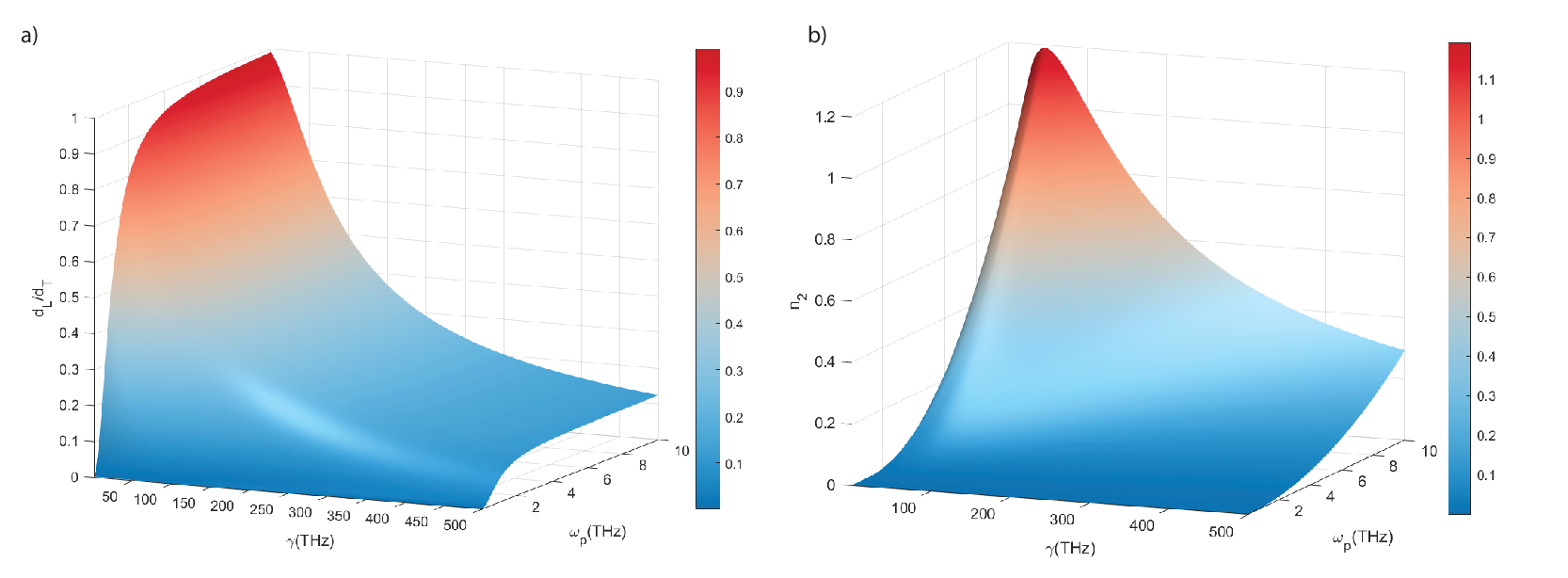}
    \caption{The special case when $\tilde{\epsilon} = \tilde{\mu}$, where fixed Lorentz model parameters of $\epsilon_\infty=1.0$ and $\omega_0=2\pi\times1.0$ THz. In (a), we plot the ratio $d_{T} / d_{TL}$ as a function of $\gamma$ and $\omega_p$. In (b), we show $n_2$ as a function of $\gamma$ and $\omega_p$.}
    \label{fig3}
\end{figure}

To further understand the $d_{T} / d_{TL}$ relation, we have also incorporated a plot of $n_2$ as a function of $\gamma$ and $\omega_p$. When the special condition $\tilde{\epsilon} = \tilde{\mu}$ is met, the reflection and transmission coefficients can be expressed as follows:

\begin{align*}
    r &= 0 \\
    t &= e^{-n_2k_0d_2}
\end{align*}

It is evident that $A = 1 - e^{-2n_2k_0d_2}$, implying that $n_2$ significantly influences $A$. Since the calculation of $d_T$ involves the integral of $A$, the value of $n_2$ also determines the limit. Figure \ref{fig3} demonstrates that an increase in $n_2$ results in a higher $d_{T} / d_{TL}$.

\section{Metamaterial Simulation Details}

We conducted simulations on three different types of metamaterial absorbers: (I) a single unit cell all-dielectric metamaterial (ADM), (II) a 2x2 supercell ADM, and (III) a single unit cell metal metamaterial (MM). The unit cell of Type I ADM is a cylindrical resonator made of silicon, while Type II supercell ADM consists of four silicon cylindrical resonators arranged in a two-by-two array. Type III MM implements a gold split ring resonator on a polyimide substrate. Figure 5 depicts the schematic for each absorber type. To validate the bandwidth to thickness limit for transparent homogeneous layers, we placed the same split ring resonator on the backside of the polyimide with a 90-degree rotation, enabling the metal metamaterial to operate in a transmissive mode while maintaining a high absorption.

\begin{figure}[ht]
    \centering
    \includegraphics[width=0.6\textwidth]{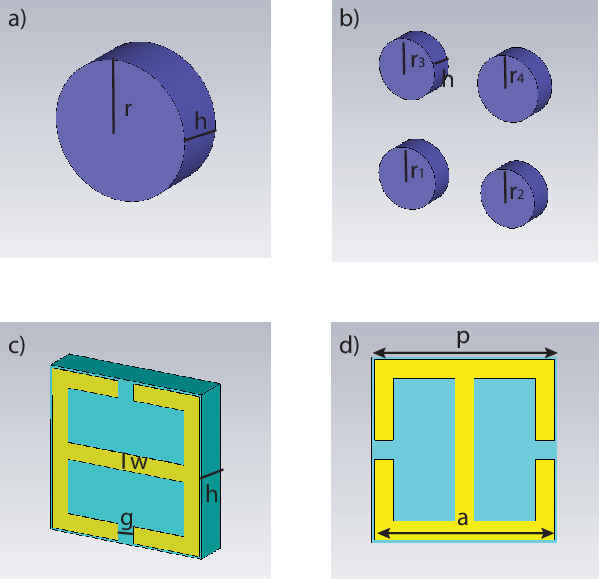}
    \caption{Schematics of Type I unit cell ADM in a), Type II supercell ADM in b), and Type III MM shown in c) \& (d).}
    \label{fig:supp_MM_schematic}
\end{figure}

The dimensions analyzed for each case are presented in Table 1. We individually adjusted the absorbers' dimensions to achieve near-unity peak absorptivity at approximately 1 THz. To achieve the desired thickness range, we scaled the entire geometry of each metamaterial absorber. We employed unit cell boundary conditions to simulate the absorption spectra of each absorber. Initially, a frequency domain solver was used for all three types of absorbers. We obtained results within a reasonable time frame for Type I and II by simulating a frequency range of 0-2 THz. However, simulating the metal metamaterial took significantly longer. Switching from the frequency domain to the time domain solver resulted in a significant speed-up in simulating metal metamaterials. To account for higher-order modes, we extended the simulation range of MM to 0-5 THz.

\begin{table*}
\centering
\caption{Geometrical parameters for metamaterials investigated for validation. Units are in $\mu$m.}
\label{MM_geom}
\begin{tabular}{|l||l|l|l|l|l|l|l|l|l|}
\hline
Type & P    & h    & $r_1$ & $r_2$ & $r_3$ & $r_4$ & w    & a    & g    \\ \hline \hline
I    & 217  & 51.4 & 61.9 & -    & -    & -    & -    & -    & -    \\ \hline
II   & 217  & 51.4 & 61.9 & 58   & 60   & 65   & -    & -    & -    \\ \hline
III  & 31.5 & 6.75 & -    & -    & -    & -    & 3.24 & 30.6 & 3.24 \\ \hline
\end{tabular}
\end{table*}

Next, we calculated the absorbers' limits. First, we retrieved the electromagnetic parameters of each metamaterial absorber \cite{smith2005electromagnetic}. Then, we calculated $\epsilon_s$ and $\mu_s$ from the low-frequency limit (0.02 THz to 0.2 THz for types I and II, and 0.05 THz to 0.5 THz for type III) of the effective material parameters obtained through parameter extraction. We calculated the integral of the absorption spectrum over the frequency range of 0 THz to 2 THz for types I and II, and 0 THz to 5 THz for type III. It should be noted that we reduced the bandwidth of the integral significantly to ensure reasonable computational costs. Hence, the large discrepancy between the limit and the metamaterial absorbers can be attributed to the much narrower bandwidth of the absorbers compared to the limit.

\bibliographystyle{unsrt}
\bibliography{main}